\documentclass[fleqn]{aa}
\usepackage{graphics}
\usepackage{amsmath,amssymb}
\usepackage{txfonts}
\usepackage{times}

\newcommand{\kms}{\,\mathrm{km}~\cdot~\mathrm{s}^{-1}}
\newcommand{\zav}[1]{\left(#1\right)}
\newcommand{\hzav}[1]{\left[#1\right]}
\newcommand{\szav}[1]{\left\{#1\right\}}
\newcommand{\Halpha}{\ensuremath{{\mathrm H}\alpha}}

\begin{document}

\title{Radiative transfer in moving media} 
\subtitle{II. Solution of the radiative transfer equation in axial
symmetry}
\titlerunning{Radiative transfer in moving media II.}

\author{Daniela Kor\v{c}\'{a}kov\'{a},
        Ji\v{r}\'{\i} Kub\'{a}t}

\authorrunning{D. Kor\v{c}\'{a}kov\'{a}, J. Kub\'{a}t}

\offprints{D. Kor\v{c}\'{a}kov\'{a}, \\
\email{kor@sunstel.asu.cas.cz}}

\institute{Astronomick\'y \'ustav, Akademie v\v{e}d \v{C}esk\'e
        republiky, CZ-251 65 Ond\v{r}ejov, Czech Republic}

\date{Received 13 September 2004 / Accepted 30 April 2005}

\abstract{
A new method for the formal solution of the 2D radiative transfer
equation in axial symmetry in the presence of arbitrary velocity fields
is presented.
The combination of long and short characteristics methods is used to 
solve the radiative transfer equation.
We include the velocity field in detail using the
Local Lorentz Transformation.
This allows us to obtain a
significantly better description of the photospheric
region, where the gradient of the global velocity is too small for the
Sobolev approximation to be valid. 
Sample test calculations for the case of a stellar wind and a
rotating atmosphere are presented.

}
 
\maketitle

%________________________________________________________________

\section{Introduction}

Radiation is the major source of information about stars.
A number of stellar properties can be obtained by comparing observed
spectra with the synthetic ones calculated from model stellar atmospheres.
A key problem in computing synthetic spectra is solving the
radiative transfer equation in stellar atmospheres, since the 
emergent radiation is formed in these regions.
There exists a large number of methods for solving the transfer
equation in a static one-dimensional case, which are, unfortunately,
inappropriate for stars with stellar winds,
rapidly rotating stars, accretion discs, and also for nebulae.
Our main interest is to study various aspects of the stellar wind in
hot stars, therefore we want to develop a method for solving the
radiative transfer equation that is well suited for this case, where the
symmetries enabling a one-dimensional approximation are broken.

There are not many methods available to solve
static multidimensional radiative transfer problems.
For optically thin regions, the Monte Carlo method 
(Boiss\'e \cite{boisse}) may be used. 
On the other hand, in the optically thick regions it is
possible to solve the transfer equation using a method that employs
the diffusion approximation (Kneer \& Heasley \cite{difuzemulti}).
Neither of these methods are suitable for stellar atmospheres, where
the transfer problem must be solved from optically thick regions to
regions where the optical thickness is very small.
The classical way of solving the radiative transfer equation in
more dimensions is the long characteristics method (Cannon \cite{dlouhy}).
This method fully describes the radiation field, but the computer time
necessary to obtain a solution may become long.
For this reason, Kunasz \& Auer (\cite{kratky}) developed the short
characteristics method, which is the best
 currently available multidimensional method.
There exist several applications of 
short characteristics methods in a Cartesian
grid (Fabiani Bendicho \cite[and references therein]{zaklad}) and
2D axially symmetric geometry (Georgiev et. al \cite{bulhar}).
Dullemond \& Turolla (\cite{dullemond}) developed a ``rotating plane''
method for axially symmetric problems based on the short
characteristics method, too. 
An efficient approach is to use adaptive grids following Folini et al.
(\cite{doris}) and Steinacker et al. (\cite{steinray}).
A review by Auer (\cite{vhled}) provides an excellent insight into the 
grounds for astrophysical multidimensional radiative transfer.

Another possibility is to apply the finite element method, which has
been recently used for multidimensional radiative transfer by, e.g.,
Richling et al. (\cite{femulti}).
However, the finite element method is not used very often for radiative
transfer in stellar atmosphere studies due to its convergence problems
caused by the about ten orders of magnitude changes of physical quantities
(e.g. opacity, density) in stellar atmospheres.
Dykema et al. (\cite{DFEIII}) tried to overcome
the instability and convergence problems using 
a modification of the finite element method, namely, the
discontinuous finite element method.
The discontinuous finite element method was also used in our previous
paper (Kor\v{c}\'akov\'a \& Kub\'at \cite{kk},
hereafter Paper I) for one-dimensional radiative transfer.

If the velocity field is present, the situation is more complicated. 
The changes in opacity and emissivity along a ray in spectral lines can
be large due to the Doppler shift, and we have to take this into
consideration.
In the continuum, we can simply use the static equation, since the
Doppler shift has only a small effect on the opacity and emissivity
coefficients. For simplicity the methods assuming monotonic velocity field
have been used very often.
The multidimensional problem with velocity fields can be
solved in the observer frame (the frame connected with the center of a
star) or in the comoving frame (the frame connected with the outflowing
material). Radiative transfer in more dimensions has usually
been solved in the observer frame (e.g. Carlsson \& Stein 
\cite{multitubingen}). 
One of the firsts works to do this is Mihalas, Auer \& Mihalas
(\cite{mihalas78}), who were able to solve a periodic velocity field in
two-dimensional planar geometry.  
Recently, van Noort et al. (\cite{vnapj,vn})
solved this problem in two dimensions.
Their code is able to cope with both spherical and cylindrical
geometry for Cartesian coordinate systems.
The observer frame is appropriate only for small velocity gradients.
The velocity difference between neighbouring depth points has
to be smaller than several times the thermal velocity.
If the velocity gradient is large, 
the Sobolev approximation (Sobolev \cite{Sobolev}) is often used.
This approximation was used for a 3D case by Folini et al. (\cite{doris}).
Both cases of small and large velocity gradients can 
be solved in the comoving frame.
For a 3D accretion disc, the radiative transfer in the comoving frame
was solved by Papkalla (\cite{papkalla}). 

However, there exists another effect that is usually neglected or taken
into account in an approximate manner, namely the rotation of stars.
Stellar rotation is usually taken into account as a convolution
of the line profile, obtained from the plane-parallel static solution,
and of the rotation profile (cf. Gray \cite{konvoluce}). 
In this technique one must make use of analytical expressions for limb
darkening, which do not give a correct description of the angular
dependence of the specific intensity.
It also does not take into account the fact that limb darkening is
strongly frequency dependent across the line profile (cf. Hadrava \&
Kub\'at \cite{hadku}).
In order to describe the radiation field and its influence on the
stellar atmosphere correctly, it is also necessary to take into account 
the Doppler shift of lines in a rotating atmosphere for the transfer of
radiation along nonradial rays. 

In this paper we present a new method to formally solve the radiative
transfer equation in axial symmetry, which does not employ the Sobolev
approximation.
Our method is based on the Local Lorentz Transformation method
(LLT) described in Paper~I for the one-dimensional case.
For studies of stellar wind, accretion discs or stellar rotation we
need a method that uses a more general geometry than
plane-parallel or spherical. However, it is not necessary to treat the whole three
dimensional space in detail. We employ axial symmetry.
The results will be more accurate than in  a plane-parallel or
spherical geometry and the calculation will be faster than for the
full 3D problem.

In the first part of this paper we present our axially symmetrical code.
The geometry of the model is described in detail,
followed by some tests of our code.
We compare these results with a spherically symmetric model atmosphere
from Kub\'{a}t (\cite{ATAmod}).
We also present some test calculations with a velocity field.
Our line profiles are compared with convolved profiles for the case of
stellar rotation.
Some results for the stellar wind are also shown.
Then we test the dependence of the computing time on the number of grid
points.
In the last part, we comment on the possibilities of the
application for our method. 

%__________________________________________________________________

\section{Method}

We solve the radiative transfer equation for the specific intensity of
radiation $I$
\begin{equation}
\vec{n}\cdot\nabla I \zav{\vec{r},\vec{n},\nu} = -
\chi\zav{\vec{r},\nu} \hzav{I\zav{\vec{r},\vec{n},\nu}-
S\zav{\vec{r},\nu}},
 \label{rpz-ob-stat}
\end{equation}
where $\chi$ is the opacity, $S$ is the source function, $\vec{n}$ is
the direction of radiation, $\vec{r}$ is the radius vector,
and $\nu$ denotes the frequency.
We assume axial symmetry and we allow for a nonzero velocity field.

The basic idea of this method is to solve
the radiative transfer problem not in the whole star, but in separated
planes intersecting the star.

\subsection{The spatial grid}\label{sitpopis}

\begin{figure}[h]
 \begin{center}
 \scalebox{.46}{\includegraphics*{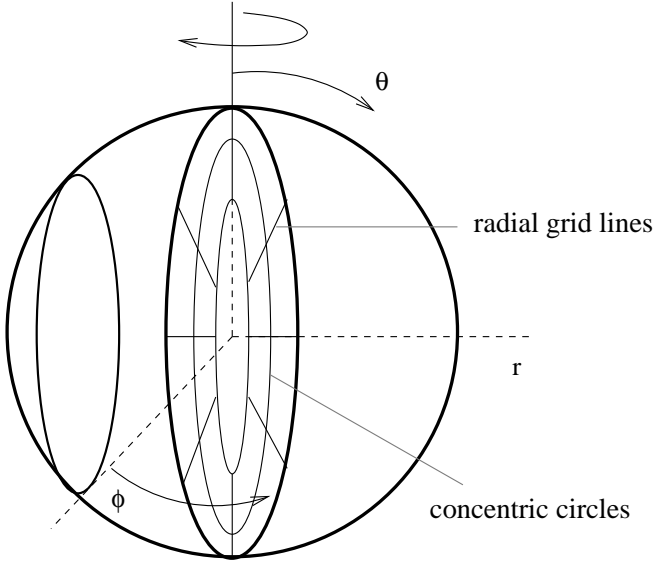}}
 \end{center}
 \caption{Coordinate system and the set of longitudinal
 planes.
 These planes are parallel to the plane $\phi=0$.
 Note that some of them intersect the opaque stellar core, where the
 diffusion approximation is valid, while the rest does not.
}
 \label{rezy} 
\end{figure}

Let us consider the spherical coordinate system ($r$, $\theta$, $\phi$).
We choose as the axis of symmetry $\theta=0$ (see Fig.\,\ref{rezy}).
First, we introduce the discretization of the radial distance $r$ and
angle $\theta$. 
Due to the axial symmetry, physical quantities do not depend on the angle
$\phi$.
The grid is chosen to give the best description of the system,
depending on the studied object.
As an example, for the study of stellar winds,
the grid of angles $\theta$ can be equidistant.
On the other hand, for accretion discs it must be finer near the
equatorial plane, where the disc resides.
In radial distance $r$ we choose the grid
(very similarly to a 1D problem)
to be equidistant in the logarithm of the radial optical depth.
A suitable choice is about 5 points per decade.
We assume the opacity and emissivity of the stellar material to be known
at the grid points.

To reduce the 3D problem to a 2D one, we do not solve the radiative
transfer equation directly in the 3D grid, but in a set of
``longitudinal planes'' intersecting the star parallel to the plane
$\phi=0$ (we ``slice the star'' -- see Fig. \ref{rezy}). 
In thin stellar atmospheres it is favourable
to choose a finer grid of longitudinal planes closer to
stellar limb, since the limb darkening is better described
by such a choice.

\begin{figure}[b]
 \begin{center}
 \scalebox{.38}{\includegraphics*{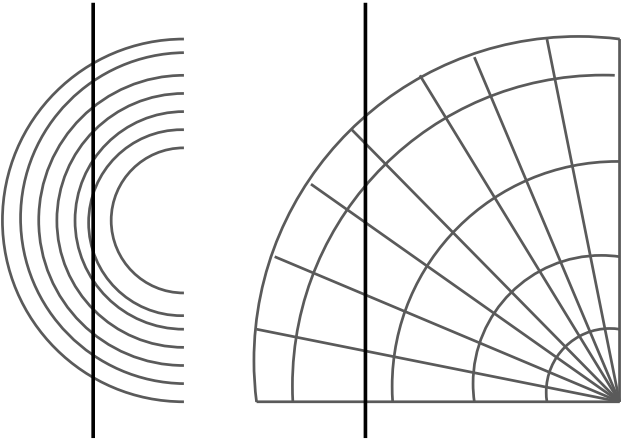}}
 \end{center}
 \caption{Choosing the grid in longitudinal planes that do not
     intersect the opaque stellar core.
     For geometrically thin (left panel) atmospheres, the grid points
     are defined using intersections of a longitudinal plane with
     concentric circles.
     For extended atmospheres (right panel), intersections of the
     longitudinal plane with radial lines are used for the definition of
     grid points.}
 \label{sitrov}
\end{figure}

The primary grid described above helps us only to interpolate the
quantities to the selected planes and finally to calculate the emergent
flux.
To solve the transfer equation we need a slightly modified grid.
For each longitudinal plane we choose the polar coordinate system and
define a grid of
concentric \emph{grid circles} and \emph{radial grid lines}
(see Fig.\,\ref{rezy}).
The grid of concentric circles corresponds to the original 3D grid in
the planes, which intersect the opaque stellar core where we do not
solve the transfer equation and where we apply the lower boundary
condition following from the diffusion approximation.
However, choosing an appropriate grid for planes which do not intersect
the opaque stellar core may become difficult, 
since the grid chosen must correspond to the geometry of the problem as well as
resolving the velocity field.
This is shown in Fig.\,\ref{sitrov}.
We must provide an appropriate grid 
for both geometrically thin and extended atmospheres.

For geometrically thin atmospheres the grid of radial distances 
in the longitudinal plane is chosen using intersection points with 
grid circles in the primary 3D grid (left panel in Fig.\,\ref{sitrov}).
For an extended atmosphere it is better to define
grid points where the plane intersects the radial grid lines
(right panel in Fig.\,\ref{sitrov}).
The reason why we need  an appropriate grid in the
planes that do not intersect the inner boundary is the necessity to
ensure sufficient spatial resolution 
for the radiative transfer equation solution.
This may be difficult in the central region of the plane.
Moreover, similar problem involves by the velocity field.
The velocity gradient in the planes far from the center can be
large, so it is necessary to have a sufficient number of grid points.

The opacity, emissivity and the source function are interpolated to the
new coordinate system.
Linear interpolation is used here, since the grid is not orthogonal.
The radiative transfer equation is solved for each longitudinal plane
independently (Section\,\ref{longplansol}).
The whole radiative field is obtained by rotating the separate planes
around the axis of symmetry (Section\,\ref{radfieldcalc}).
 
%======================================================================

\subsection{The frequency grid}

The frequency grid is to be chosen to enable the most efficient
description of the radiation field.
We set the frequency interval using both the largest line width and the
total Doppler shift caused by the global motion.
For example, for Doppler profile  this interval is
\begin{align}
 \nonumber
 \nu \in
\langle
& 
  \nu_{0} \left(1-(\text{v}_{\infty}+\text{v}_{rot_{0}})/c \right)-
      2 \Delta \nu_{D},  \\ 
  & \nu_{0} \left(1+(\text{v}_{\infty}+\text{v}_{rot_{0}})/c \right)+
      2 \Delta \nu_{D}
\rangle,
 \label{interval}
\end{align}
where $\text{v}_{\infty}$ is the terminal velocity, $\text{v}_{rot_{0}}$ 
the rotation velocity in photosphere.
$\Delta \nu_{D}$ denotes the broadest Doppler halfwidth,
which corresponds to the highest temperature. 
For very extended atmospheres or planetary nebulae
(optically thin medium) we must
take into account that the regions with $-\text{v}_{\infty}$ and  
$\text{v}_{\infty}$ ``see'' each other. In that case, we must 
multiply  $\text{v}_{\infty}$ in Eq. \eqref{interval} by two.
We take the frequency step to be equidistant for the case 
of the Doppler line profile. This step must be small enough
to describe the line profile variations in the region with
low temperature. The step is determined in the following way.
We first  typically set 5 points per line equidistantly at the depth
point with the narrowest line profile (which is usually the depth point
with the lowest temperature).
Then we cover the whole frequency interval \eqref{interval} with
frequency points using this frequency step.
This frequency grid is then used in other depth points where the line
profile is broader.
For a Voigt profile we need to extend the frequency interval
\eqref{interval}, but the frequency step in the extended part of the
interval may be larger.

%======================================================================
\subsection{Solution in longitudinal planes}
\label{longplansol}

In the given longitudinal plane we
introduce the polar coordinate system using
grid circles and radial grid lines
defined in the preceding subsection (\ref{sitpopis}).
The values of temperature, density and velocity are known 
at the grid points. First, we solve the transfer equation in the plane
starting at the outer boundary.
Once the radiation field in the direction towards the center 
is known, we can continue to solve the transfer equation 
in the opposite direction,
i.e. from the inner boundary towards the outer one.
At the inner boundary the diffusion approximation may be
taken for planes intersecting the opaque stellar center
as a boundary condition.
In other planes (that do not intersect the stellar center), we adopt as
the lower boundary condition the intensity taken from the previously
calculated solution from the outer boundary inwards.

%.......................................................................
\subsubsection{The solution from the outer boundary to the central
regions}

\begin{figure}[h]
 \begin{center}
 \scalebox{.4}{\includegraphics*{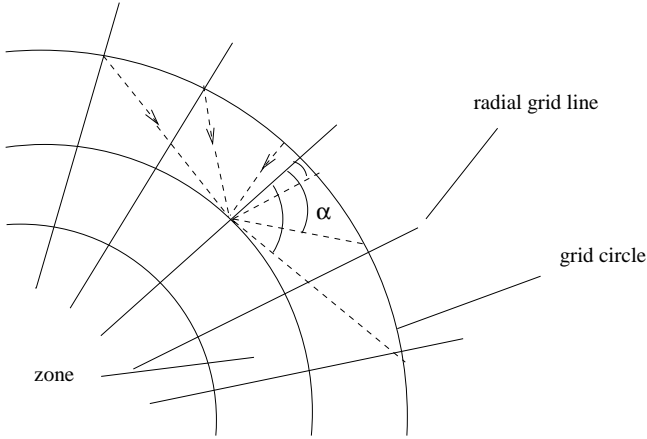}}
 \end{center}
 \caption{Diagram for the solution of the radiative transfer
 equation in the
 longitudinal plane starting at the upper boundary to the central
 regions.
 The rays along which we perform the formal solution are plotted using
 the dashed line.
}
 \label{dolucast}
\end{figure}

We begin the calculations at the outer boundary (stellar surface), where
the boundary condition (i.e. the incoming intensity $I$) is known. 
At each inner grid point we choose several rays per quadrant (see
Fig.\,\ref{dolucast}). The rays start at the outer grid circle and
end at a given grid point.
Along these rays we solve the transfer equation.
The angle distribution of these rays may be the same as
often used in the case of the
plane-parallel geometry, where the angle cosines
$\mu=\cos \alpha$ ($\alpha$ is the angle between the ray and
the radial direction, see Fig. \ref{dolucast})
are chosen to be the roots of Legendre polynomials in the interval $(0,1)$
($\mu = 0.8872983346$, $0.5$, $0.1127016654$)
to ensure better numerical accuracy of the angle integration
(Press et. al \cite[section 4.5]{recipes}). Three 
rays at a given point is usually sufficient, because the whole
radiation field is
obtained by summing the information from all longitudinal planes
intersecting the given point (see Fig.\,\ref{celkpole}).  
However, in some situations it is better to use more (up to 9) rays
per quadrant to overcome a numerical error introduced by the necessary
interpolation of intensity described below.
As one can see in Fig.\,\ref{dolucast}, the rays
start at the preceding grid circle. This means that the rays
do not usually start at a grid point and that they
may also intersect some radial grid lines.

\begin{figure}[h]
 \begin{center}
  \scalebox{.48}{\includegraphics*{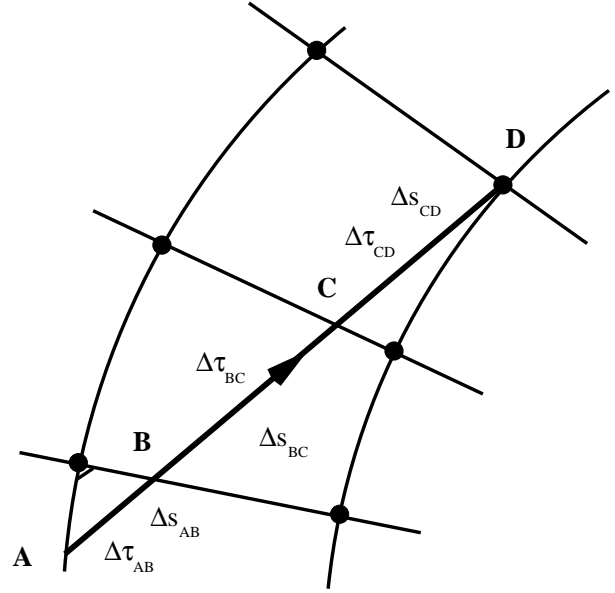}} \hspace{4mm}
 \end{center}
 \caption{The scheme for the integration along the ray
for a determination of the specific intensity at the point $D$.
The specific intensity at the grid points of outer grid circle is known.
The value of the specific intensity at the point $A$ is determined
by linear interpolation from values at neighbouring grid points.
The values of the source function and opacity in the points $B$ and $C$
are obtained by linear interpolation.}
 \label{integrace}
\end{figure}

The solution diagram is illustrated in Fig.\,\ref{integrace}.
We perform a linear interpolation of the source function and opacity to
obtain their values at points $A$, $B$, and $C$ and of the incoming
intensity for the value at point $A$.
The optical depth difference $\Delta \tau$ is calculated along the ray
between the individual intersection points
(abscissas $AB$, $BC$, and $CD$) using the linear approximation,
\begin{equation}
\label{opthloubka}
 \Delta \tau_{(AB)} = \frac{\chi_{A}+\chi_{B}}{2}
 \Delta s_{(AB)},
\end{equation}
and similarly for $\Delta \tau_{(BC)}$ and $\Delta \tau_{(CD)}$.
Here $\chi_{A}$ and $\chi_{B}$ are opacities at respective points
and $\Delta s_{(AB)}$ is the geometrical distance between points
$A$ and $B$.

We solve the equation of radiative transfer by parts between all
intersection points along the ray.
We assume a linear dependence of the source function on the optical
depth between the intersection points.
For the interval $AB$ the solution is
\begin{equation}\label{rovnice} 
 I_{(B)}=I_{(A)} e^{-\Delta \tau_{(AB)}}+ \int^{\Delta \tau_{(AB)}}_{0}
       S(t) e^{[-(\Delta \tau_{(AB)}-t)]} dt,
\end{equation}
and similarly for intervals $BC$ and $CD$.
The final intensity $I_{(D)}$ at the grid point $D$ is thus
determined by three successive applications of the equation
\eqref{rovnice}.
In this manner we obtain the specific intensity in the downward
solution at every grid point.  
It may, of course, happen that the geometric distance along the ray in
the given cell is very small (the ray near the point $B$ in the
Fig.\,\ref{integrace}). In this case
we do not solve the transfer problem in this cell and we simply set
$I_{(B)}=I_{(A)}$.
Doing this, we eliminate a numerical instability. 
Usually, it is enough if the condition 
$|\cos (\pi /2- \theta - \alpha)|> 10^{-7}$ ($\alpha$ has
the same meaning as in Fig.\,\ref{dolucast}) is fulfilled, even if
it is much more accurate to estimate it using the optical depth difference. 

There are several reasons why we prefer only a linear dependence of the
source function on other higher order approximations.
First, it is numerically stable and second, it is much faster. 
The higher order optical depth interpolation may sometimes lead
to numerical errors by adding new extrema
(cf. Auer \cite{interpolace}).
They may become significant especially in moving medium. 
There is only one reason why linear interpolation should not be used:
because the diffusion approximation in deep optically thick layers
would be inaccurate.
Since we do not solve the transfer problem using the short
characteristics method, this inaccuracy is not large in our case.
Our characteristics are longer and the transfer equation is solved in
several steps, so the information from the farther regions is naturally
included. Therefore we chose the safer linear interpolation.

%.......................................................................
\subsubsection{The solution from the central regions to the outer
boundary}

\begin{figure}[h!]
 \begin{center}
  \scalebox{.5}{\includegraphics*{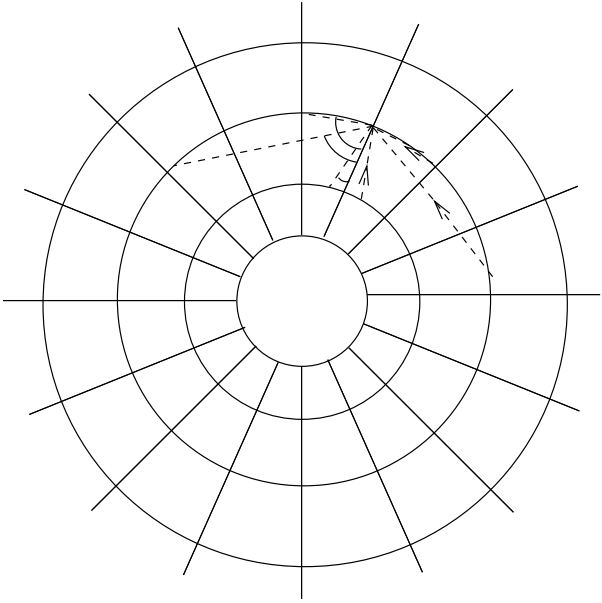}} 
 \end{center}
 \caption{Diagram for the solution
   of the radiative transfer equation from the center of the star to the
   outer boundary.
   The rays along which we solve the formal solution are plotted using
   the dashed line.
 }
 \label{nahoru}
\end{figure}

The upward solution is very similar to the previous step.
The procedure is depicted in Fig.\,\ref{nahoru}.
We lead the rays to the grid points 
using the same angles as in the previous case.
The rays start either at the preceding grid circle (closer to the
center) or at the same circle as the grid point.
At the intersection points we interpolate the opacity and the source
function. Between these points we calculate the optical depth
using \eqref{opthloubka} and solve the transfer equation
using \eqref{rovnice} as in the previous case of downward integration.

\begin{figure}[h]
 \begin{center}
  \scalebox{.56}{\includegraphics*{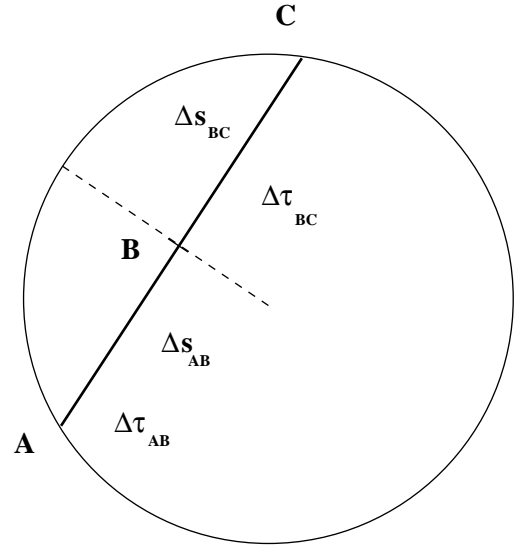}} 
 \end{center}
 \caption{Diagram for a solution of the radiative transfer equation in
 the central region in the planes, which do not intersect the region of
 validity of the diffusion lower boundary condition.}
 \label{stred}
\end{figure}

From Fig.\,\ref{rezy}, one can see that there exist planes that
do not intersect the inner boundary region.
For these planes we adopt the intensity calculated from the previous
step (solution from the upper boundary to the stellar center) as the
lower boundary condition.
However, the solution of the transfer equation in the central grid
circle must be performed with care.
The situation is shown in the Fig.\,\ref{stred}.
First, we solve the radiative transfer equation from the intersection
point $A$ to the center of the abscissa $AC$ 
(point $B$) and then to point $C$. 
The physical quantities at point $B$ we obtain by linear
interpolation from points $A$ and $C$.
The optical depth difference is calculated using \eqref{opthloubka} and
the specific intensity is determined using \eqref{rovnice}.
The value of the intensity at point $C$ determines
the lower boundary condition for further solutions outwards.
For planes, which intersect the region of validity of the diffusion
approximation (the stellar core), the situation is easier.
We simply use the appropriate lower boundary condition.

%-----------------------------------------------------------------------
\subsection{The full radiation field} \label{radfieldcalc}

\begin{figure}[h]
 \begin{center}
 \scalebox{.46}{\includegraphics*{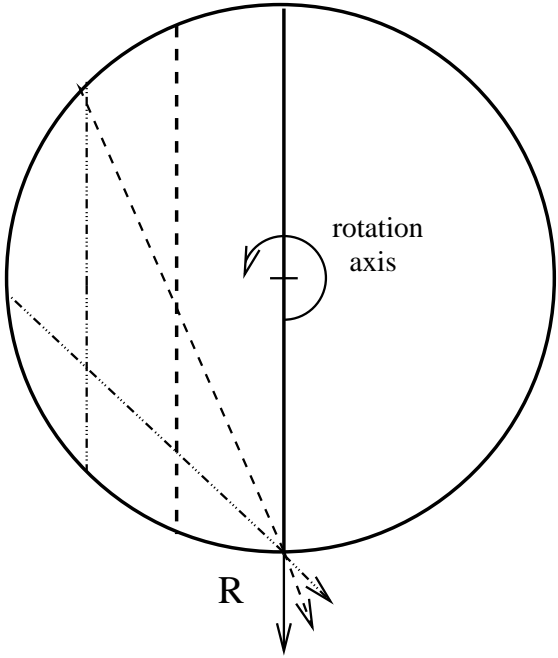}}
 \end{center}
 \caption{
   Diagram describing the determination of
   the whole radiation field.
   The rotation axis for 
   longitudinal
   planes is perpendicular to the page.
   Longitudinal planes (projected as lines in this figure) are rotated
   along the axis to intersect at a point $R$.
   Due to the axial symmetry, the radiation field for \emph{all}
   directions is obtained by a rotation of all
   longitudinal planes around the axis of symmetry.
 }
 \label{celkpole}
\end{figure}

To obtain the whole radiation field we take the advantage of the
symmetry of the problem. 
We know the radiation field at grid points in all directions lying
inside the longitudinal planes.
The radiation field in the whole star is then obtained by rotating 
the longitudinal planes around the axis of symmetry $\theta=0$ 
(see\,Fig.\,\ref{celkpole}).
This gives a sufficient description of the specific intensity.
The emergent radiation flux towards the observer
is calculated by integrating specific emergent
intensity over the stellar disc.

%-----------------------------------------------------------------------
\subsection{Velocity field}

The radiative transfer equation in moving media can be solved either 
in the observer frame or in the comoving frame.
Since the coefficients of emissivity and opacity are angle dependent
in the observer frame, we solve this problem rather in the comoving
frame. This frame, which is coupled with the moving medium, 
is generally a non-inertial frame.
The form of the radiative transfer equation in the comoving frame
we can obtain either from the general relativity or from the special
relativity by assuming a set of Local Lorentz Transformations. 

In the Paper I we introduced an
LLT (Local Lorentz Transformation)
method for a solution of the radiative transfer equation in
one-dimensional moving media based on an application of Lorentz
transformations on cell boundaries and a static radiative transfer
equation solution between them.
The method was tested by a comparison with an exact solution of the
radiative transfer equation using the discontinuous finite element
method (see Fig.\,8 in Paper I), which solves the comoving transfer
equation including the $\partial / \partial \nu$ 
(Doppler) and
$\partial / \partial \mu$ (aberration) terms.
Here we apply the same idea for the case of a two-dimensional
arbitrarily moving medium.

\begin{figure}[h]
  \scalebox{.34}{
      \includegraphics*{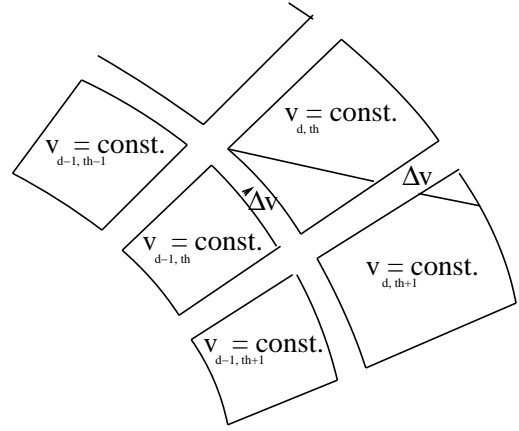}}
 \caption{
   The scheme for the inclusion of the velocity field. Velocity
   within each cell is constant  and a change of velocity is possible
   only at the cell boundaries. 
 } 
 \label{osarych}
\end{figure}

In our method, we replace the solution
of the radiative transfer equation in the comoving frame
by a set of Local Lorentz Transformations and a solution of the transfer
equation in corresponding local inertial frames.
In these inertial frames we solve a static radiative transfer equation,
since it is Lorentz invariant.
To do this, we must consider the constant velocity of
every cell and the change of the velocity we allow only at the
cell boundary (see Fig. \ref{osarych}).
Two other conditions must be fulfilled there.
First, velocity with respect to the observer frame
must be low to neglect aberration. 
This permits us to solve the radiative transfer along straight 
characteristic lines.
This assumption is valid, if 
$\text{v}/c \ll 0.01$ (Mihalas et al. \cite{CMF3},
Hauschildt et al. \cite{H95}).
Second, let us assume a steady-state fluid. Due to this assumption 
we can neglect the time delay between different parts of
the medium. The correctness of this approach
has been discussed in more detail in Paper I,
where this method was tested for a plane-parallel geometry.

The basic solution scheme within cells, which was described
in Section\,\ref{longplansol},
is not affected by the presence of the velocity field.
We project the velocity field to the longitudinal planes.
At cell boundaries we perform the Lorentz transformation of frequency
$\nu$
(see also Eq.\,(24) in paper I)
\begin{equation}
 \nu^\prime = \nu \sqrt{1 - \frac{\zav{\Delta \text{v}}^2}{c^2}}
 \zav{1-\frac{\vec{n}\cdot\Delta\vec{ \textbf{v}}}{c}}
\end{equation}
($\Delta\vec{ \textbf{v}}$ is the velocity difference between cells).
Since the velocity field
is not relativistic, we can simply use the classical Doppler law, which
speeds up the solution.
The Lorentz transformation of intensity
$I^\prime \zav{\nu^\prime} = I \zav{\nu} \zav{{\nu^{\prime 3}}/{\nu^3}}$
(see Eq,\,(23) in Paper I)
at the cell boundary has a
negligible effect, and we do not take it into account, since the
intensity is proportional to the third power of the frequency ratio. 
This is close to one in most stellar applications.
Since the equation of the radiative transfer is Lorentz invariant,
we solve the static equation of the radiative transfer 
\eqref{rovnice} within each cell.

Since we solve the transfer equation in the comoving frame, we have
to do one additional step.
We know the source function and opacity in the frequency grid points in
the rest frame coupled with a given cell.
The incoming intensity from a neighbouring cell is known at different
frequencies, which are Doppler shifted according to the velocity
difference between the cells. This means that
we must interpolate the intensity to frequency points coupled to
the rest frame of a cell. 

To ensure a correct treatment of line transfer we have to
guarantee that the frequency shift
at the cell boundaries due to Doppler effect
is smaller than a quarter of a Doppler halfwidth (see Paper~I).
If it is not, we must make the grid finer.
This condition must be fulfilled even if we calculate with the Voigt
profile since the center of the line has a Doppler core. 
This is the only limitation of this method.
If the velocity gradient
is too high, we must refine the grid, which
slows down the calculation.

%%%%%%%%%%%%%%%%%%%%%%%%%%%%%%%%%%%%%%%%%%%%%%%%%%%%%%%%%%%%%%%%%%%%%%%%

\section{Test calculations}

Tests of the method are based on a model atmosphere of
a main sequence B star with
effective temperature $T_\mathrm{eff}= 17\times 10^{3}\ \mathrm{K}$,
gravitation acceleration $\log g=4.12$,
and radius $3.26 R_{\odot}$. 
We obtain the state parameters, electron density and temperature, 
using the hydrostatic spherically symmetric model atmosphere code ATA 
(Kub\'{a}t \cite{ATAmod}).
We consider no incoming radiation
as the outer boundary condition and the diffusion approximation as
a lower boundary condition.

%=======================================================================
\subsection{Static case}

\begin{figure}[h] 
 \scalebox{.66}{\includegraphics{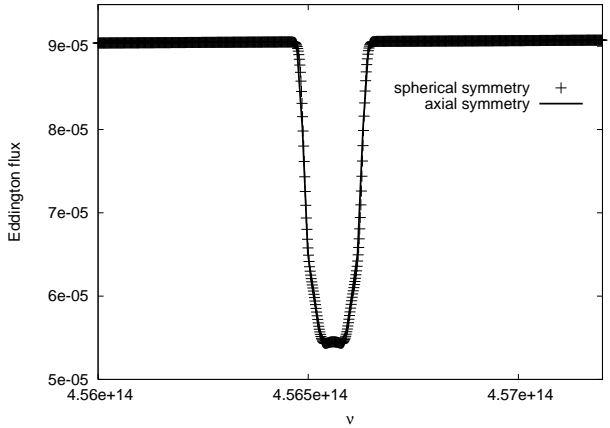}}
 \caption{The comparison of Eddington flux in the {\Halpha} line from
    the static spherically symmetric code ATA (crosses) and the axially
    symmetric one (solid line).}
 \label{jirkaja}
\end{figure}

For a basic comparison we took a model atmosphere
calculated using the static computer code ATA.
The radiative transfer in ATA is solved using the long
characteristics method
and Feautrier variables (see Kub\'at \cite{ATAdis,ATA1,ATAmod}).
We compare the result from the ATA code with the flux obtained from our
new code (Fig.\,\ref{jirkaja}).  
The line profile was chosen to be Doppler here and the input parameters
in the axially symmetric code are independent of $\theta$.
This comparison for
this simplest case proves the basic correctness of the new code.

\subsubsection{Limb darkening}

\begin{figure}[h] 
 \scalebox{.7}{\includegraphics{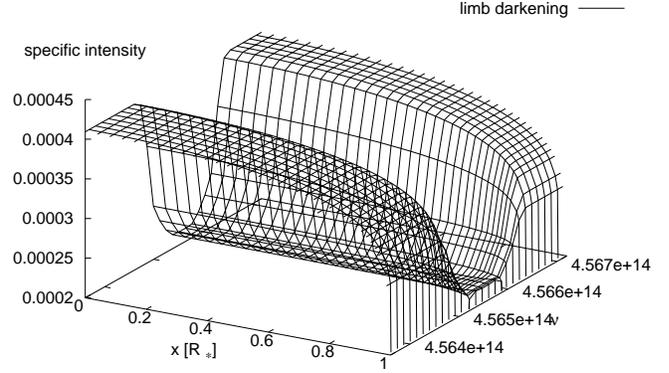}}
 \caption{
  Theoretical limb darkening obtained using our code across
  a half of the stellar disk in the {\Halpha} line.
  $R_\star$ denotes the stellar radius.
 }
 \label{okraj} 
\end{figure}

\begin{figure}[h]  
 \scalebox{.66}{\includegraphics{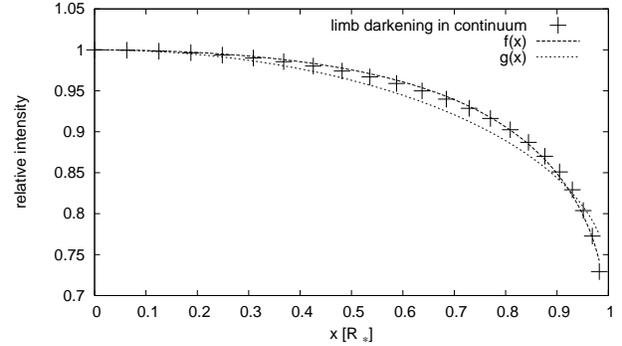}}
 \caption{Limb darkening in the continuum.
   Crosses denote data computed from our model.
   The fitting functions $f(x)$ and $g(x)$ which express the commonly
   used limb darkening laws are defined by equations \eqref{fxkont} and
   \eqref{gxkont}.  }
 \label{okrajkontinuum} 
\end{figure}

\begin{figure}[h]  
 \scalebox{.66}{\includegraphics{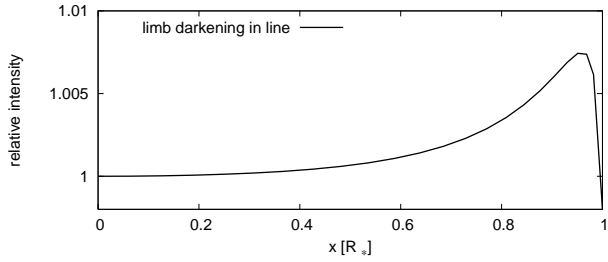}}
 \caption{
  Center-to-limb variations at the central frequency
  of the {\Halpha} line.
  Note the apparent weak limb brightening before final darkening at the
  stellar limb.}
 \label{okrajcara} 
\end{figure}

An important result arising
from the solution of the transfer equation is the
limb darkening law. 
In Fig.\,\ref{okraj} we plotted the limb darkening
law across the {\Halpha} line profile calculated from our code.
We choose as an x-axis the distance from the center of star in
  star radius units
($x=1$ for $r=R_\star$, where $R_\star$ is the stellar radius).
At this scale the effect of limb
  darkening is more clearly visible.
To show this effect in more detail, we extracted the dependence of the
specific intensity on the distance from the center of the stellar disc
for a continuum frequency (Fig.\,\ref{okrajkontinuum}) and for the
central frequency of the {\Halpha} line (Fig.\,\ref{okrajcara}). 
The obtained data for the continuum are compared with
functions usually used to express the limb darkening law.
The first function ($f(x)$) is adopted from Allen's Astrophysical
Quantities (\cite{allen}),
\begin{equation}
 \label{fxkont}
  I(x)=1-a-b+a (1-x^{2})^{1/2}+b(1-x^{2}) \equiv f(x).
\end{equation} 
Using the least squares method we obtain the parameters $a=0.55 \pm
0.02$ and $b=-0.20 \pm 0.01$ to fit our results to 
Eq.\,\eqref{fxkont}.
Limb darkening is also often described by the
law (Gray \cite{konvoluce}),
\begin{equation}
 I(x)=(1-\epsilon)+\epsilon  (1-x^{2})^{1/2} \equiv g(x).
 \label{gxkont}
\end{equation}
Fitting this case we obtain the value of the parameter
$\epsilon= 0.277 \pm 0.008$.

Figure \ref{okrajcara} shows the limb darkening in the center of
the {\Halpha} line.
As one can see, brightening instead of darkening is observed near 
the stellar limb, which corresponds to the effect of 
flash spectra in solar chromosphere.
The same result was obtained also by Hadrava and Kub\'{a}t
(\cite{hadku}).  

%=======================================================================

\subsection{Stellar wind}

The case of the moving atmosphere is first tested for a spherically
symmetric stellar wind, for which we adopt the 
beta law
for the dependence of the wind velocity $\text{v}$ on radius $r$
(see, e.g.,
Lamers \& Cassinelli \cite{CL99})
\begin{eqnarray}
 \text{v}(r)=\text{v}_\infty \szav{1-\hzav{1-
 \zav{\frac{\text{v}_{R}}{\text{v}_\infty}}^{\frac{1}{\beta}}} \frac{R}{r}}^{\beta}.
 \label{beta}
\end{eqnarray} 
We choose a velocity in the photosphere of 
$\text{v}_{R}= 200\kms$ and a terminal velocity $\text{v}_{\infty}=2000\kms$. 

\begin{figure}[h]
 \scalebox{.38}{\includegraphics{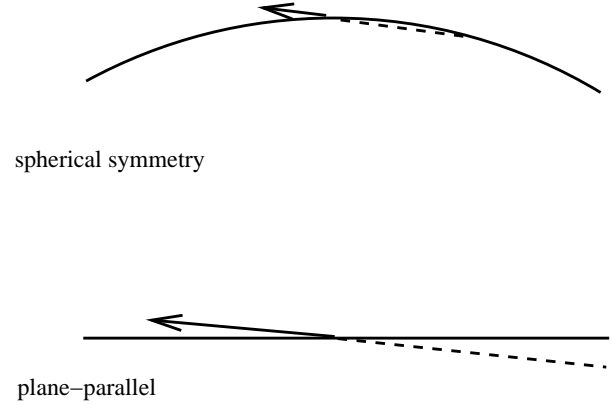}}
 \caption{
   Intensity obtained from plane-parallel and axially symmetric geometries
   differ especially for very sideways (almost tangent) rays.
 }
 \label{osa-lte-obr}
\end{figure}

To check the consistency of our new method we compare it 
with the plane-parallel method described in Paper I. 
The plane-parallel radiative transfer equation is solved
using the discontinuous finite element method.
The velocity field is included using the Local Lorentz Transformation
similarly to this paper. In Paper I
the Local Lorentz Transformation method was compared with the
solution of the radiative transfer equation
using the discontinuous finite element method,
where the frequency
term was consistently included into the solution as an independent
variable. Since the velocity field is non-relativistic, calculations with
aberration give the same results as without it (in agreement with
Mihalas et al. \cite{CMF3}, Hauschildt et al. \cite{H95}).
\begin{figure}[h]
 \scalebox{.66}{\includegraphics{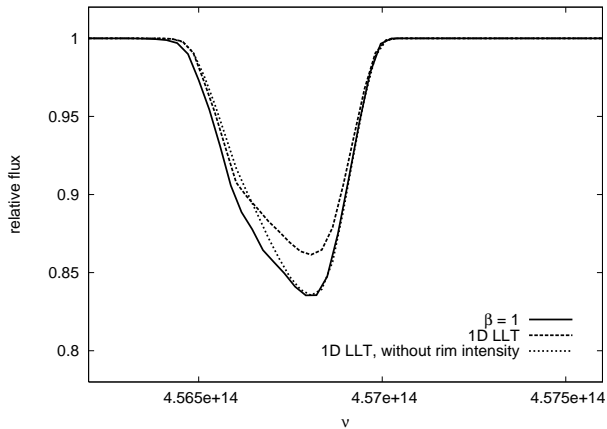}}
 \caption{ 
   The comparison of the $\Halpha$ line profile obtained from our
   axially symmetric model (solid line) and a plane-parallel model
  (denoted by 1D\,LLT).
   The dotted line is obtained using the same model, but
   the radiation with $\mu<0.26$ is
   not taken into account to approximate the curvature of the atmospheric
   layers (see also Fig.\,\ref{osa-lte-obr}).
   The beta velocity law parameter $\beta=1$ for all profiles.
 }
 \label{osa-lte}
\end{figure}
The result from plane-parallel geometry is plotted using the dashed line
in Fig.\,\ref{osa-lte}. As we can see,
the agreement with the new (axially symmetric
-- solid line) results is not good.
The reason is in the geometry used.
Rays with large angle (low $\mu$) in a curved atmosphere (spherically
symmetric or axially symmetric) do not even touch regions that are
optically thick in continuum, whereas for a plane parallel geometry
\emph{all} rays with $\mu>0$ finally reach the continuum optically thick
regions (see Fig.\,\ref{osa-lte-obr}).
As a consequence, there is more continuum radiation in a plane parallel
geometry, which results in shallower line profiles.
This has been well known for a long time (see, e.g., Kunasz et al.
\cite{spher2}).
This is why we plot another line profile, which also is obtained  using
the plane-parallel geometry, but in which the intensity incoming from
directions with $\mu<0.26$ is set to zero.
We can see a very good agreement in the blue wing of the
line profile, where we see central parts of the
star and where the assumption of plane-parallel atmosphere is acceptable.
However, the sphericity effects, limb darkening and limb
brightening influence the red wing of the line.
A different profile from a axially symmetric code is in agreement with the
 fact the line profile obtained in spherical geometry is different
from that obtained in a plane parallel geometry.
The difference depends on the sphericity effects on the temperature
structure (see, e.g., Kunasz et al. 1975, Gruschinske \& Kudritzki 1979,
Kubat 1995, 1999). 

\begin{figure}[h]
 \scalebox{.66}{\includegraphics{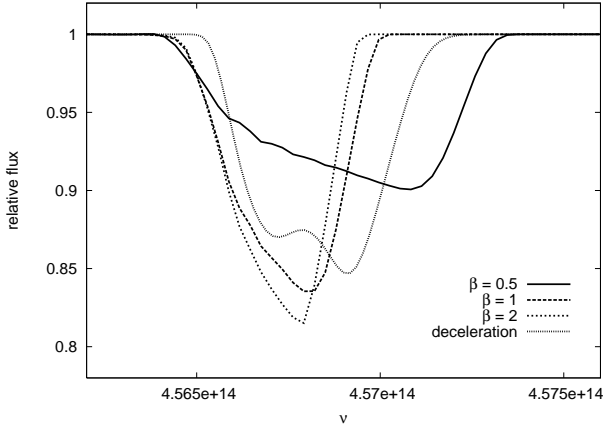}}
 \caption{The profile of the {\Halpha} line for the case of a
 stellar wind for three values of parameter $\beta=0.5, 1, 2$ 
 (see Eq.\,\eqref{beta}).
 For comparison, the line profile  affected by decelerating velocity field
 is also plotted (fine dotted line).}
 \label{vitrobr} 
\end{figure}

In  Fig.\,\ref{vitrobr} we compare the line profiles 
for three different values of the parameter $\beta$ ($0.5$, $1$ and $2 $).
To show the possibility of our code to solve a more general
velocity field we plot in Fig.\,\ref{vitrobr} the line profile obtained
using a decelerating velocity field. 
We choose a linear dependence of the velocity on the radial distance
in a logarithmic scale. The photospheric and terminal velocities
are 2000 $\kms$ and 200 $\kms$, respectively.

The most important point
is the velocity gradient in the region of line formation.
For this reason, lines corresponding to different values of the
$\beta$ (in Fig. \ref{vitrobr}) parameter have different profiles,
even if the values of the photospheric and infinity velocity are
the same. We do not see the P Cygni profile, since
the input parameters are based on the hydrostatic model,
which produces a geometrically thin atmosphere.
To obtain a P Cygni profile, not only enough high velocity gradient, but
a sufficiently extended line formation region are necessary.
Consequently, only a blue shifted absorption profile appears
in our case. 
Note that similar line profile shapes were obtained by Noerdlinger
\& Rybicki (\cite{ppvel}) for a plane-parallel expanding atmosphere.

%=======================================================================
\subsection{Stellar rotation}
\label{rotace}

To test the stellar rotation
we assume a spherical star and
we consider the power-law dependence of the
rotation velocity on the radius,
\begin{eqnarray}\label{rotzakon}
 \text{v}(r)=\text{v}(R_{*})\left( \frac{r}{R_{*}} \right)^{-j},
 \label{rotacevr}
\end{eqnarray}
where $R_{*}$ is the stellar radius and $\text{v}(R_{*})$
is the rotation velocity in the photosphere. 
We choose the parameter $j$ to be equal to $1$, which corresponds to
equatorial discs formed by a stellar wind (Kroll \& Hanuschik
\cite{kroll}) using conservation of the angular momentum.
Stars which rotate near their critical velocity are far from
being spherically symmetric.
For example, the ratio of the polar and equatorial radii of
$\alpha$~Eridani was determined to be about one half (Domiciano de Souza
\cite{achernar})
or even $1/5$ (Jackson et al. \cite{ach}).
In these stars the approximation of spherical symmetry fails and we must
solve the transfer problem in a more general geometry.
However, although gravity darkening is neglected in our test case, 
our code is able to easily handle this effect.

We compare
line profiles calculated using our model with a rotational velocity
field and detailed radiative transfer  
with other possibilities of calculation of the rotationally
broadened profile.

The standard and most commonly used way is to solve a detailed static
plane-parallel radiative transfer equation for a static atmosphere
and to convolve the resulting profile with a rotation profile using 
the relation (see Gray \cite{konvoluce}),
\begin{eqnarray}
\frac{F_{\nu}}{F_{c}}=H(\nu)*G(\nu)=
\int^{\infty}_{-\infty} H(\nu-\Delta \nu) \ G(\Delta \nu) \ d\Delta \nu.
\label{konvolucerov}
\end{eqnarray}
Here, $F_{\nu}$ is the flux for a given frequency,
$F_{c}$ is the continuum flux, $H(\nu)$ is the normalized
 flux from the nonrotating star.
The rotation profile $G(\nu)$ is equal to
\begin{multline}
   G(\nu)=
     \frac{2(1-\epsilon)
     \sqrt{1- \left( \frac{\Delta \nu}{\Delta \nu_{\max}} \right)^{2}} +
              \frac{1}{2} \pi \epsilon 
     \sqrt{1- \left( \frac{\Delta \nu}{\Delta \nu_{\max}} \right)^{2}}}
          {\Delta \nu_{\max} \pi \left(1-\frac{\epsilon}{3}\right)} 
 \label{rotacniprofil}
\end{multline}
in the interval $| \Delta \nu | < \Delta \nu_{\max}$, and $G(\nu)=0$
elsewhere.
This expression assumes the limb darkening law in the form
\eqref{gxkont} and the parameter $\epsilon$ is the same as
in the Eq.\,\eqref{gxkont}.

A more exact and also computationally more expensive approach also
uses the emergent radiation from the static atmosphere, but takes into
account the angle dependence of the specific intensity.
The resulting profile is then calculated by integrating
the specific intensity across the stellar disc. 
This approach will be called ``integrated static profile'' hereafter.
In these two latter approaches the radius dependence of the rotational
velocity \eqref{rotacevr} cannot be taken into account 
and the photosphere is tacitly assumed to rotate as a rigid body.

\begin{figure}[h] 
 \scalebox{.66}{\includegraphics{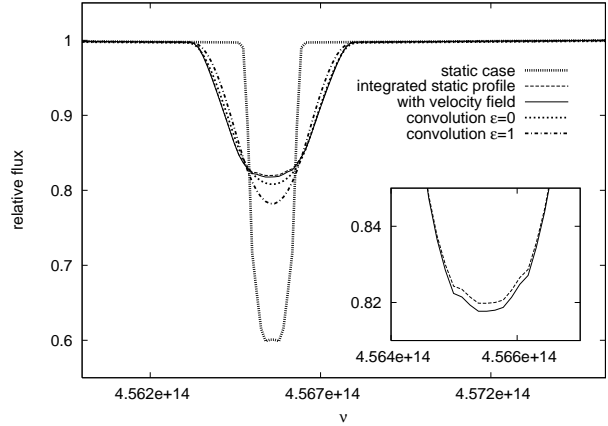}}
 \caption{The comparison of the static {\Halpha} line profile and
   rotation profiles obtained using different methods
   for a rotation velocity $0.2$ of the critical rotation velocity.
   The line profile calculated exactly using our 2D radiative transfer
   code with a corresponding velocity field
   \eqref{rotacevr} is plotted using the full line.
   The line profile that takes into account the surface velocity field,
   but with the local line profiles those calculated from the solution of
   the static atmosphere, is plotted using the dashed line
   (integrated static profile in the figure).
   Line profiles calculated using the convolution of a static profile and
   the rotation profile with $\epsilon=0$ and $\epsilon=1$ are plotted
   using the dotted line and dashed-dotted line, respectively.
   In addition, the profile calculated for the static case (no rotation)
   is plotted using the thick dotted line.
    }
 \label{rotaceobr} 
\end{figure}

The most exact solution is to calculate the emergent radiation using
a full solution of the radiative transfer equation
in a moving atmosphere, as has been done using our method.
Results for the rotation velocity
described by Eq.\,\eqref{rotzakon} with
$\text{v}(R_{*})=0.2 \text{v}_{c}$
are shown in Fig.\,\ref{rotaceobr}.
Here, $\text{v}_{c}=\sqrt{{G M_{*}}/{R_{*}}}$
is the critical rotational velocity, which is equal to $540\kms$ for our
case.
In this figure we plotted the profile of the {\Halpha} line in the
static case, the profile obtained from convolving the static profile
with the rotation profile \eqref{rotacniprofil}
using values $\epsilon=0$ and $1$, the one from the
integration of the static profile over the stellar disc, and the
``true'' profile calculated with the full influence of the rotation
velocity field.
The ratio of the equivalent width in the static case and the flux
obtained from integrating of the line profile over the disc is of the
order of $10^{-5}$, which represents good numerical accuracy. 

Both line profiles obtained using convolution \eqref{konvolucerov} are 
deeper than those calculated from our code.
The difference between line depth obtained by convolution and line depth
obtained from our code
$(I_\mathrm{our}-I)/(1-I_\mathrm{our})$
is about $5$\% for parameter 
$\epsilon=0$ and almost $20$\% for $\epsilon=1$.
The error of five percent is not too large,
since sometimes the observed spectra have a lower accuracy.
However, for high S/N spectra one may have an accuracy better than $1\%$,
which makes the error of $5\%$ significant.
There is no doubt about the detectability of a $20\%$ difference.
The commonly used value of the parameter, $\epsilon \sim 0.6$, yields
remarkable differences as well. This difference
is a consequence of the dependence of limb darkening on frequency
across the line profile.
As we can see from Fig.\,\ref{okraj}, the continuum intensity decreases
with increasing distance from the center of the stellar disk.
On the other hand, the intensity in the center of the line is
significantly changed very close to the limb (see
Fig.\,\ref{okrajcara}).
These effects cannot be described by the simple
formulae \eqref{fxkont} and \eqref{gxkont}.
The difference between the line profile calculated by integrating the
specific intensity over the disc and the profile which
includes the velocity field and
detailed radiative transfer is very small
(see the magnified part of Fig.\,\ref{rotaceobr}), as expected,
since the radial velocity gradient is relatively small
and the line forming region is very thin.
For more extended and rapidly rotating
sources these effects will be amplified.

%=======================================================================
\subsection{Accretion discs}

Our method is also appropriate for solving the radiative transfer
equation in accretion discs.
It can take advantage of the accretion disc geometry.
Since the disc is densest and, consequently, optically thickest
close to the equatorial plane, we can employ
the possibility of unevenly distributed angles $\theta$ in the
definition of the primary grid. 

In addition to having a better resolution of the disc, the chosen grid
is also able to describe possible jets.
Using this method we can also include in the calculations the boundary
region, winds from this region and from the inner disc, as well as the
hot corona beyond the disc.
Another advantage of this method is the ability to handle both
optically thin and thick discs. 
Emergent spectra from the accretion disk of a cataclysmic variable
calculated using our method were presented in Kor\v{c}\'akov\'a et al.
(\cite{kkkcv}).

%=======================================================================

\subsection{The tests of the grid}

\begin{figure} 
 \scalebox{.66}{\includegraphics{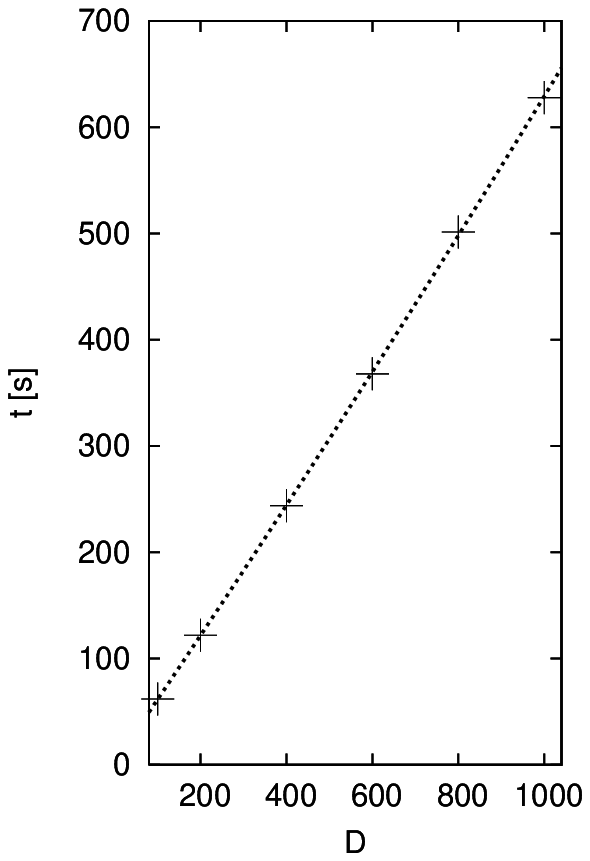}}
 \scalebox{.66}{\includegraphics{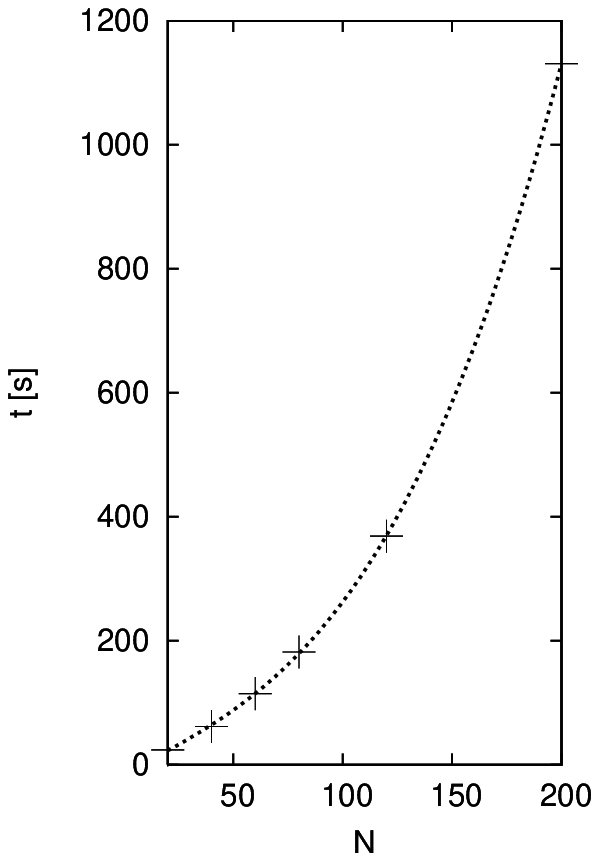}}
 \caption{The dependence of the computing time on the number of depth
   points (\textit{left panel}) and on the number of frequency points 
   (\textit{right panel}).}
 \label{sit}
\end{figure}

\begin{figure}
 \scalebox{.7}{\includegraphics{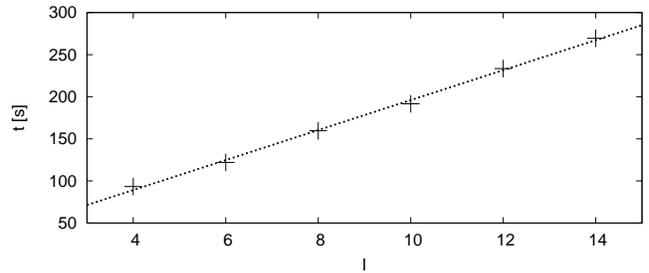}}
 \caption{ The dependence of the computing time on the number of 
   angular grid points.}
 \label{sitI}
\end{figure}

The grid tests show a linear dependence of the computing time on the
number of geometrical depth points $D$ (see Fig.~\ref{sit}, left panel).
The right panel of the same figure shows the dependence of the computing
time on the number of frequency points $N$.
The fitting function is a polynomial of the third order.
 Although one may expect only linear dependence of the computing
time on a number of frequency points, higher order dependence
is due to the necessity of interpolating the intensity
for each frequency point at each boundary of the cells
due to the Doppler shift. 
In Fig.~\ref{sitI} we show the dependence of computing time on
the number of angular grid points $I$.
The fitting function is also a straight line. 
 
%%%%%%%%%%%%%%%%%%%%%%%%%%%%%%%%%%%%%%%%%%%%%%%%%%%%%%%%%%%%%%%%%%%%%%%%

\section{Conclusions}

We presented a new method for solving the radiative transfer
equation in axial symmetry with the possibility of including
arbitrary velocity fields.
The basic idea is to solve the transfer equation in planes, that
intersect the object.
In a given plane a combination of long and short characteristics
methods is used.
This method allows us to better describe the global character of the
radiation field and the resulting computing time is not too long.
The velocity field is taken into account using the Lorentz
transformation of frequency, which allows us to solve the transfer
problem in the region with a small velocity gradient as well as for high
velocity gradients.
This technique is very useful for studying stellar wind, where it is
applicable to the stellar wind region together with the
stellar photosphere.

Tests of this method were performed for a model atmosphere of a
B type star with $T_\mathrm{eff}= 17\times 10^{3}\ \mathrm{K}$,
gravitational acceleration $\log g=4.12$, and radius $3.26 R_{\odot}$.
For a static spherically symmetric atmosphere our code gives
correct results, as can be seen from Figure\,\ref{jirkaja}.
We also present the limb darkening law for our model (Fig.\,\ref{okraj}).
Note that the line profile shows limb brightening at the central
line frequency (Fig.\,\ref{okrajcara}).
This result is in agreement with the results of Hadrava and Kub\'{a}t
(\cite{hadku}).

Further tests were performed in the presence of a velocity field.
For an expanding atmosphere (stellar wind) we adopt the
classical $\beta$-velocity law (\ref{beta}).
The resulting line profile is shown in Fig.\,\ref{vitrobr}.
Since we take the input parameters (temperature, density) from the
hydrostatic code, we do not obtain a P~Cygni profile.
However, the line profile is shifted to the blue part and deformed.
Note that from these blue parts of ultraviolet lines one can determine
the wind terminal velocity.

In section \ref{rotace} we studied the application of our
method to the problem of stellar rotation.
Fig.~\ref{rotaceobr} shows the necessity of including the
frequency dependence of limb darkening in calculating 
rotationally broadened line profiles.
On the other hand, the difference between the flux calculated by
integrating the Doppler shifted static profile across the disc and the
``true'' flux that included the solution of the transfer problem with a
velocity field is very small in a thin atmosphere. 

The dependence of computational time on the number of angular points
(Fig.~\ref{sitI}) as well as on the number of geometrical points is
linear (Fig.~\ref{sit}, left panel).
The dependence on the number of frequency points is more complicated,
which stems from the interpolation of intensity due to the Doppler shift
at cell boundaries.
 
The limitation of this method is only in an axially symmetrical
  approach and not very steep velocity gradient. In
the latter case it is necessary to refine the grid and the computing time
becomes longer.
It is not possible to use it for the relativistic velocities too, since
the aberration effect must be included in this case. 

Our method for solving the radiative transfer equation is especially suitable
for including  stellar winds, since it is able to handle
the outer region of a stellar wind together with the stellar
photosphere.
This is necessary, above all, for line-driven winds of hot stars.
The process of initiating this type of wind near the photosphere is not
fully understood yet.
Calculations that involve the wind together with the quiet and possibly
almost static stellar atmosphere may be able to resolve this problem in
the future.
Our method can also be used to solve the transfer problem in
very rapidly rotating stars.
Gravitational darkening is too large in these objects, so it is not
possible to neglect the dependence of physical properties on the angle and
to calculate using the assumption of spherical symmetry.
The results of the application of our method can become useful for
interpreting interferometric observations.
The geometrical flexibility of our method is also very useful for studying
extremely nonspherical objects such a accretion discs.
It is possible to simultaneously include  the disc, central object, jets
and hot corona.
Another advantage of this method is the possibility of solving 
both optically thin and thick discs.

\begin{acknowledgements}
The authors would like to thank the referee for valuable comments on
the manuscript and Dr. Ad\'ela Kawka for her comments.
This research has made use of NASA's Astrophysics Data System.
This work was supported by grants GA \v{C}R 205/01/1267 and
205/04/P224.
The Astronomical Institute Ond\v{r}ejov is supported by projects
K2043105 and Z1003909. 
\end{acknowledgements}

%----------------------------------------------------------------------

\newcommand{\SAM}[1]{in {\em Stellar Atmosphere Modeling}, I.~Hubeny,
        D.~Mihalas \& K.~Werner eds., ASP Conf. Ser. Vol. 288, p.~#1}

%--------------------------

\begin{thebibliography}{}

\bibitem[1963]{allen} Allen, C. W. 1963, {\em Astrophysical Quantities},
  University of London, The Athlone Press

\bibitem[2003a]{interpolace} Auer, L. 2003a, \SAM{3}

\bibitem[2003b]{vhled} Auer, L. 2003b, \SAM{405}

\bibitem[1990]{boisse} Boiss\'{e}, P. 1990, A\&A, 228, 483

\bibitem[1970]{dlouhy} Cannon, C. J. 1970, ApJ, 161, 255  

\bibitem[2003]{multitubingen} Carlsson, M., \& Stein, R. F. 2003,
        \SAM{505}

\bibitem[2003]{achernar} Domiciano de Souza, A., Kervella, P., Jankov,
        S., et al. 2003, A\&A, 407, L47

\bibitem[2000]{dullemond} Dullemond, C. P., \& Turolla, R. 2000, A\&A,
        360, 1187

\bibitem[1996]{DFEIII} Dykema, P. G., Klein, R. I., \& Castor, J. I.
        1996, ApJ, 457, 892

\bibitem[2003]{zaklad} Fabiani Bendicho, P. 2003, \SAM{419}

\bibitem[2003]{doris} Folini, D., Walder, R., Psarros, M., \& Desboeufs,
        A. C. 2003, \SAM{433}

\bibitem[2003]{bulhar} Georgiev, L. N., \& Hillier, D. J. 2003,
        \SAM{437}

\bibitem[1976]{konvoluce} Gray, D. F. 1976, {\em Observation and
	Analysis of Stellar Photospheres}, John Wiley \& Sons, New York

\bibitem[1979]{} Gruschinske J., \& Kudritzki R.-P. 1979, A\&A, 77, 341

\bibitem[2003]{hadku} Hadrava, P., \& Kub\'at, J. 2003, \SAM{149}

\bibitem[1995]{H95} Hauschildt, P. H., Starrfield, S., Shore, S.,
        Allard, F., \& Baron, E. 1995, ApJ, 447, 829

\bibitem[2004]{ach} Jackson, S., MacGregor, K. B., \& Skumanich, A.
	2004, ApJ, 606, 1196

\bibitem[1979]{difuzemulti} Kneer, F., \& Heasley, J. N. 1979, A\&A, 79,
        14

\bibitem[2003]{dankadis} Kor\v{c}\'akov\'a, D. 2003, PhD thesis, Masaryk
	University Brno

\bibitem[2003]{kk} Kor\v{c}\'akov\'a, D., \& Kub\'at J. 2003, A\&A, 401,
        419
%Ku12:
	(Paper I)
%

\bibitem[2004a]{kkkpop} Kor\v{c}\'akov\'a, D., Kub\'at J., Krti\v{c}ka,
	J., \& \v{S}lechta, M. 2004a, in {\em The A-Star Puzzle}, IAU
	Symp.  224, J. Zverko,
	J. \v{Z}i\v{z}\v{n}ovsk\'y, S. J. Adelman \& W. W. Weiss eds.,
	Cambridge, Univ. Press, p. 533

\bibitem[2004b]{kkkcv} Kor\v{c}\'akov\'a, D., Kub\'at J., \& Kawka, A.
	2004b, in {\em 14th European Workshop on White Dwarfs},
	D. Koester \& S. Moehler eds., ASP Conf. Ser., submitted

\bibitem[1997]{kroll} Kroll, P., \& Hanuschik, R. W. 1997, in: IAU Coll.
        163, 494 

\bibitem[2003]{krttue} Krti\v{c}ka, J. 2003, \SAM{259}

\bibitem[2004]{krtecek} Krti\v{c}ka, J., \& Kub\' at, J. 2004, A\&A, 417,
        1003

\bibitem[1993]{ATAdis} Kub\'at, J. 1993, PhD thesis, Astronomick\'y
	\'ustav AV \v{C}R Ond\v{r}ejov

\bibitem[1994]{ATA1} Kub\'at, J. 1994, A\&A, 287, 179

\bibitem[1995]{SPWD} Kub\'at, J. 1995, A\&A 299, 803

\bibitem[1999]{SPAGB} Kub\'at, J. 1999, NewA 4, 157

\bibitem[2003]{ATAmod} Kub\'at, J. 2003, in {\em Modelling of Stellar
        Atmospheres}, IAU Symp. 210, N. E. Piskunov, W. W. Weiss \&
        D. F. Gray eds., ASP, A8

\bibitem[1988]{kratky} Kunasz, P., \& Auer, L. H. 1988, JQSRT, 39, 67

\bibitem[1975]{spher2} Kunasz, P. B., Hummer, D. G., \& Mihalas, D.
	1975, ApJ, 202, 92

\bibitem[1999]{CL99} Lamers, H. J. G. L. M., \& Cassinelli, J. P. 1999,
	{\em Introduction to Stellar Winds}, Cambridge Univ. Press,
	Cambridge

\bibitem[1978]{Mihalas} Mihalas, D., 1978, {\em Stellar Atmospheres},
	2nd ed., W. H. Freeman \& Comp., San Francisco

\bibitem[1976]{CMF3} Mihalas, D., Kunasz, P. B., \& Hummer, D. G. 1976,
        ApJ, 206, 515

\bibitem[1978]{mihalas78} Mihalas, D., Auer, L. H., \& Mihalas, B. R.
        1978, ApJ, 220, 1001

\bibitem[1974]{ppvel} Noerdlinger, P. D., Rybicki, G. B. 1974, ApJ, 193,
	651

\bibitem[1995]{papkalla} Papkalla, R. 1995, A\&A, 295, 551

\bibitem[1986]{recipes} Press, W. H., Flannery, B. P., Teukolsky, S. A.,
      	\& Vetterling, W. T. 1986, {\em Numerical Recipes, The Art of
        Scientific Computing}, Cambridge Univ. Press, Cambridge

\bibitem[2001]{femulti} Richling, S., Meink\"{o}hn, E., Kryzhevoi, N.,
        \& Kanschat, G. 2001, A\&A, 380, 776

\bibitem[1947]{Sobolev} Sobolev, V. 1946, {\em Dvizhushchiesia obolochki
	zvedz}, Leningr. Gos. Univ., Leningrad

\bibitem[2003]{steinray} Steinacker, J., Henning, T., Bacmann, A., \&
        Semenov, D. 2003, A\&A, 401, 405

\bibitem[2002]{vnapj} van Noort, M., Hubeny, I., \& Lanz, T. 2002, ApJ,
        568, 1066

\bibitem[2003]{vn} van Noort, M., Hubeny, I., \& Lanz, T. 2003,
        \SAM{445}


\end{thebibliography}
\end{document}